%% file: IPDPS24-alya-vectorized.tex
\documentclass[fleqn,10pt]{SelfArx} 

\input{00-preamble.tex}

\JournalInfo{} 
\Archive{} 

\PaperTitle{Exploiting long vectors with a CFD code:\\ a co-design show case}

\Authors{\
Marc Blancafort\textsuperscript{1}*,
Roger Ferrer\textsuperscript{1},
Guillaume Houzeaux\textsuperscript{1},
Marta Garcia-Gasulla\textsuperscript{1},
Filippo Mantovani\textsuperscript{1}} 
\affiliation{\textsuperscript{1}\textit{Barcelona Supercomputing Center Barcelona, Spain}} 
\affiliation{*\textbf{Corresponding author}: marc.blancafort@bsc.es} 

\Keywords{co-design, vectorization, CFD, RISC--V, optimization, performance, portability}
\newcommand{\keywordname}{Keywords} 

\input{01-abstract.tex}   

\begin{document}

\maketitle 

\tableofcontents 

\thispagestyle{empty} 

\input{10-intro.tex}            
\input{30-background.tex}       
\input{38-methodology.tex}      
\input{50-implementation.tex}   
\input{60-evaluation.tex}       
\input{20-related-work.tex}     
\input{70-conclusions.tex}      

\section*{Acknowledgment}

Supported by the EuroHPC Joint Undertaking (JU): FPA N.~800928 (EPI), SGA N.~101036168 (EPI-SGA2), and GA N.~101093393 (CEEC CoE). The JU receives support from the EU Horizon 2020 research and innovation programme and from Croatia, France, Germany, Greece, Italy, Netherlands, Portugal, Spain, Sweden, Denmark and Switzerland. The EPI-SGA2 project, PCI2022-132935 is also co-funded by MCIN/AEI /10.13039/501100011033 and by the UE NextGenerationEU/PRTR.

\bibliographystyle{IEEEtran}
\bibliography{99-sdv}

\end{document}

%% file: 00-preamble.tex
\usepackage{lineno}
\modulolinenumbers[5]
\usepackage[utf8]{inputenc}
\usepackage[T1]{fontenc}
\usepackage[english]{babel} 
\usepackage{cite}
\usepackage{amsmath,amssymb,amsfonts,mathrsfs}
\usepackage{algorithmic}
\usepackage{graphicx}
\usepackage{array}
\usepackage[caption=false,font=footnotesize,labelfont=sf,textfont=sf]{subfig}
\usepackage{textcomp}
\usepackage{booktabs} 
\usepackage{multirow}
\usepackage{listings}
\newcolumntype{M}[1]{>{\centering\arraybackslash}m{#1}}
\newcolumntype{N}{@{}m{0pt}@{}}
\def\BibTeX{{\rm B\kern-.05em{\sc i\kern-.025em b}\kern-.08em
    T\kern-.1667em\lower.7ex\hbox{E}\kern-.125emX}}

\newcommand{\blind}[1]{
\ifdefined\BLIND
{\color{gray}(omitted for blind review)}%
\else
{#1}%
\fi
}

\usepackage{xspace}
\newcommand{\ie}{i.e.,\xspace}
\newcommand{\eg}{e.g.,\xspace}

\newcommand{\mn}{MareNostrum~4\xspace}

\newcommand{\vs}{\texttt{VECTOR\_SIZE}\xspace}
\xspaceaddexceptions{=}

\definecolor{color1}{RGB}{0,0,90} 
\definecolor{color2}{RGB}{0,20,20} 
\definecolor{gray}{RGB}{0,0,90} 

\usepackage{hyperref} 

\hypersetup{
	hidelinks,
	colorlinks,
	breaklinks=true,
	urlcolor=color2,
	citecolor=color1,
	linkcolor=color1,
	bookmarksopen=false,
	pdftitle={Title},
	pdfauthor={Author},
}

\usepackage{url}

\usepackage{listings} 
\definecolor{col1}{HTML}{1E88E5}
\definecolor{col2}{HTML}{D81B60}
\definecolor{col3}{HTML}{43A047}
\definecolor{col4}{HTML}{F4511E}
\definecolor{col5}{HTML}{E259FF}

\usepackage{pifont}
\RequirePackage[normalem]{ulem}
\RequirePackage{color}\definecolor{RED}{rgb}{1,0,0}\definecolor{BLUE}{rgb}{0,0,1}

\providecommand{\DIFdel}[1]{{\protect\color{RED}\sout{#1}}}
\providecommand{\DIFdel}[1]{}



%% file: 01-abstract.tex
\Abstract{
A current trend in HPC systems is the utilization of architectures with SIMD or vector extensions to exploit data parallelism.
There are several ways to take advantage of such modern vector architectures, each with a different impact on the code and its portability. For example, the use of intrinsics, guided vectorization via pragmas, or compiler autovectorization. Our objectives are to maximize vectorization efficiency and minimize code specialization.
To achieve these objectives, we rely on compiler autovectorization. We leverage a set of hardware and software tools that allow us to analyze in detail where autovectorization is suboptimal. Thus, we apply an iterative methodology that allows us to incrementally improve the efficient use of the underlying hardware.
In this paper, we apply this methodology to a CFD production code. We evaluate the performance on an innovative configurable platform powered by a RISC-V core coupled with a wide vector unit capable of operating with up to 256 double precision elements.
Following the vectorization process, we demonstrate a single-core speedup of 7.6$\times$ compared to its scalar implementation. Furthermore, we show that code portability is not compromised, as our solution continues to exhibit performance benefits, or at the very least, no drawbacks, on other HPC architectures such as Intel x86 and NEC SX-Aurora.
}

%% file: 10-intro.tex
\section{Introduction}\label{secIntro}

After decades of an oligopoly among providers and architectures powering the world's most powerful supercomputers, the landscape has recently become more heterogeneous. The increasing demand for computing power has led to two significant developments: the emergence of discrete accelerators, such as GPUs, and the diversification of CPUs. This diversification includes the incorporation of SIMD and vector units, exploring design concepts that were previously only theoretical and largely unexplored from a technological standpoint.

Furthermore, the business model has transitioned away from monolithic companies that define their own ISA, design and implement their own CPUs, and eventually produce their chips. Instead, there is a shift toward a model where IP providers collaborate with implementers. For example, in the case of Fugaku, Arm and Fujitsu have partnered to achieve this. An even more ``radical'' example is RISC-V, which offers its ISA as royalty-free, allowing anyone to implement it without charges. The concept behind RISC-V is to establish a standard ISA upon which companies can base their CPU and accelerator implementations.

All this generates a potential diversity of computational platforms that could appear in modern data centers with relatively short notice. The question so far has been: ``What can we do to efficiently use new architectures to come?''. The question that drives this paper dares to go one step further: ``What can we do to prepare applications to be portable and performant and even influence the adoption of new technologies in data centers?''.

Our attention is primarily directed towards CPUs, particularly those that are SIMD and vector accelerated. In addition to Intel's SIMD extensions, we are witnessing the emergence of various elegant and diverse vector architectures. These range from Arm's Scalable Vector Extension~\cite{rico_arm_2017} to the NEC SX-Aurora accelerator~\cite{yamada2018vector,gomez2021efficiently,takahashi_performance_2023}.
The already mentioned open ISA RISC-V also proposes a vector extension and several research projects have been looking at it as a possible candidate for efficient HPC compute nodes~\cite{perotti2022new,cavalcante2023spatz}.
The open nature of the RISC-V approach coupled with the availability of relatively mature research platforms shaped our research question and the idea of co-design that we would like to present in this work.
Having software and hardware tools that allows software development and early test on new architectures allows in fact not only to prepare complex HPC codes to be ready to run efficiently before the actual hardware is available but also to influence the development of the system software (\eg compilers) and even provide feedback to the hardware engineers that design the new systems.

This paper would like to present the experience of preparing a CFD code to efficiently run on a long vector architecture by leveraging the autovectorization capabilities of the compiler.
We believe that our experience can work as best practice for other scientists approaching new architectures with complex code.
Beside showing the iterative steps for improving performance of a complex HPC code leveraging a RISC-V prototype implementing a vector unit, we show that the code changes proposed guarantee code portability and allow performance improvement also on state-of-the-art already existent HPC platforms.
CFD codes are responsible for a large fraction of computing hours in modern data centers and this is the reason why we choose one of those codes for our evaluation.


The main contributions of this paper are:
{\em i)} we apply an iterative method to implement a fast co-design cycle involving a RISC-V vector prototype;
{\em ii)} we present the optimization of a complex CFD code to exploit compiler autovectorization capabilities that drives to a speedup up to 7.6$\times$;
{\em iii)} we show that performance portability still holds and even improves on other HPC platforms.
%

The rest of this paper is structured as follows:
Section~\ref{secBackground} introduces the concepts and the technological component involved in our study.
Section~\ref{secMethodology} summarizes the methodology and the vision behind the steps of the implementation.
Section~\ref{secImplementation} details the implementation of the optimizations performed on the CFD code and the quantification of the performance benefits on the RISC-V vector prototype.
In Section~\ref{secEvaluation} we provide details about the evaluation, reporting performance measurements on different HPC machines.
Section~\ref{secRelatedWork} presents the state-of-the-art studies at the best of the knowledge of the authors.
We conclude the paper with our findings and comments in Section~\ref{secConclusions}.

%% file: 30-background.tex
\section{Background}\label{secBackground}

In this section, we describe the technological background needed for the understanding of our study.
We present the RISC-V prototype platform and the tools that allow us to experiment with the iterative co-design steps.
A special subsection is dedicated to the metrics that we use to quantify the benefits of the vectorization.
We also present Alya, the CFD code that we selected for our vectorization study.
To complete the background, we briefly introduce two HPC systems used for evaluating the results of our effort and prove that the resulting code is portable without performance issues.


\subsection{RISC-V vector prototype}\label{secRiscVProto} 

RISC-V VEC is a design developed by \blind{the European Processor Initiative project (EPI)}. 
It is a RISC-V-based CPU implementing a RISC-V scalar core developed by \blind{SemiDynamics} coupled with \blind{Vitruvius}, a Vector Processing Unit (VPU) designed and implemented by \blind{BSC~\cite{vitruvius-bsc}}.
The RISC-V vector (RVV) specification does not constrain the size of the vector registers, leaving it up to the implementor.
Moreover, RVV uses vector-length agnostic (VLA) programming model, enabling the same binary run under any machine implementing the RISC-V vector extension, no matter the size of vector registers.
 
\blind{Vitruvius} follows version 0.7.1 of the RVV extension standard and its vector length is 16~kbits: each register can hold up to 256 double-precision elements.
Peak performance is reached when all 8 lanes, each one including a Floating Point Unit (FPU) designed by \blind{the University of Zagreb}, operate over the registers. This delivers an FMA throughput of 32 cycles, \ie 16~GFLOPS per core operating at 1~GHz.

While the actual chip is being produced, the \blind{EPI} team made available a set of tools called \blind{``Software Development Vehicles'' (SDV)} that allows software development to experiment, measure and optimize the performance of generic codes.
The workflow to study a code using those tools is presented in \blind{\cite{sdv-codesign}}.
These tools include a compiler, performance tools, a software emulator and an FPGA emulator:

\subsubsection{Compiler}\label{secCompiler}

We leveraged the LLVM-based \blind{EPI} compiler with support for vector instructions.
We can introduce vector instructions via guided auto-vectorization, pragmas, or builtins.
To maintain the portability across other vector machines as well as ensure performance portability on non-accelerated platforms, we have adopted the guided auto-vectorization option, thus not making the code dependent on any system.
Table \ref{tabFlagsVec} shows the flags used to compile the mini-app.

\begin{table}[htbp]
\renewcommand{\arraystretch}{1.2}
  \centering
  \rowcolors{2}{gray!15}{white}
  \caption{Compiler options used for enabling auto-vectorization.}
  \label{tabFlagsVec}
  \resizebox{\columnwidth}{!}{%
  \begin{tabular}[\textwidth]{M{4.1cm}M{4.1cm}}
    \rowcolor{gray!30}
    \textbf{Flag}                         & \textbf{Description} \\
    -O3                                   & Set highest level of compiler optimization \\
    -ffp-contract=fast                    & Allows floating-point expression contracting such as FMA \\
    -m\blind{epi}                         & Enable auto-vectorizer \\
    -mcpu=\blind{avispado}                & Enable specific instruction code generator \\
    -combiner-store-merging=0             & Avoids inefficient combinations of memory operations \\
    -vectorizer-use-vp-strided-load-store & Allows the vectorizer to use strided vector memory accesses \\
    -disable-loop-idiom-memcpy            & Disable transforming loops into memcpy \\
    -disable-loop-idiom-memset            & Disable transforming loops into memset \\
  \end{tabular}%
}
\end{table}

\subsubsection{RISC-V vector emulator}\label{secVehave}

%
The RISC-V vector emulator used for this study is called \blind{Vehave}. 
It allows executing applications with vector instructions in machines that do not natively support the RVV extension by catching the vector illegal instructions and emulating them by software.
Immediately after, the RISC-V vector emulator returns the execution to the main process.
This software emulator enables us to run vector applications on commercial RISC-V platforms and accelerate the software development, as we can run vector applications on scalar RISC-V systems with higher frequencies while emulating a vector environment.

\subsubsection{FPGA prototype}\label{secFpgaSDV}

While the actual chip of the RISC-V VEC is being produced, an FPGA implementation is made available for software development.
While the architectural features of the system emulated on the FPGA are the same of the RISC-V VEC chip, the number of cores is scaled down.
At the moment of the study the FPGA prototype offers a single RISC-V scalar core coupled with a VPU, a memory subsystem of 1~MB of L2 data cache and 4~GB of DDR4 main memory.
The bitstream of such design is mapped inside a Virtex UltraScale+ HBM VCU128 PCIe FPGA board\footnote{\url{https://www.xilinx.com/products/boards-and-kits/vcu128.html}} and operates at 50~MHz.

\subsubsection{Performance Tools}\label{secPerformanceTools}

Since our work is based both on an emulator and a real implementation of the RISC-V VEC system, we need tools that allow us to gather performance metrics and guide our optimization steps.
On the one hand, we used \blind{Extrae}, a generic instrumentation tool able to gather data from the PAPI library during the execution of applications on different computing platforms and store it on a trace file.
This data is used to compare the speed-ups when the program is running on a real RISC-V Vector (RVV) platform.
On the other hand, the RISC-V vector emulator offers support for tracing the vector instructions executed.
It gathers information such as the vector length (VL) or vector instruction type.
The data can be later re-arranged into a trace-friendly format.
Both traces can be visualized using \blind{Paraver} to showcase the potential bottlenecks and suggest code optimizations in a visual-friendly format.

\subsection{Performance Metrics}\label{secPerformanceMetrics}

Since the design and the compiler/system software is under development, we defined a set of rigorous and representative performance metrics to quantify the exploitation of vector architecture capabilities.

When we execute an application we use hardware counters to have a dynamic and quantitative observation of the status of the application.
In our work we focus on the total cycles $c_t$ and the cycles spent to execute vector instructions (including accesses to the memory), $c_v$.
A similar notation can be used for referring to the instructions executed: we call $i_t$ the total number of instruction executed and $i_v$ the number of vector instructions.
Other important quantities to monitor are the data cache misses that we indicate with $m_{L1}$ or $m_{L2}$ depending on which level of cache we are referring to, and the vector length (or VL) of an instruction, that we simply indicate with $vl$.
Since we analyze different phases of the code, when needed, we use an exponent to highlight to which phase we are referring to.

\begin{figure}[!htbp]
\centering
\includegraphics[width=\columnwidth]{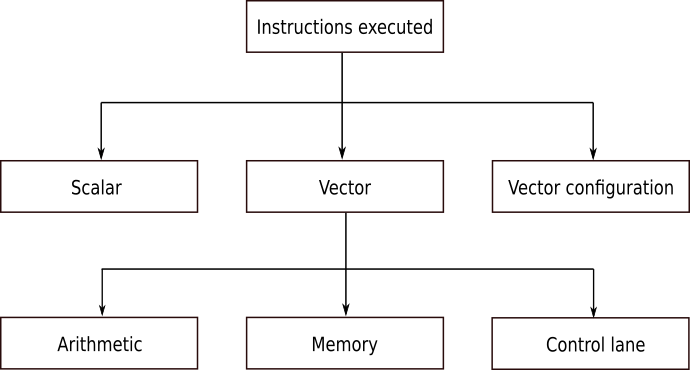}
\caption{Instruction hierarchy.}
\label{figInstructionHierarchy}
\end{figure}

Based on this assumptions, we can define the following metrics:

\begin{itemize}

\item $\displaystyle M_v = {i_v}/{i_t}$ represents the {\bf vector instruction mix}. It is a value that ranges between 0 and 1 and tells us how much our code has been vectorized, \eg if $M_v=0$ after a compilation with the auto-vectorizer activated, tells us to check the compiler output because no vector instructions have been emitted.
\item $\displaystyle A_v = {c_v}/{c_{t}}$ represents the {\bf vector activity}. It ranges as well between 0 and 1 and tells us how much time/cycles have been spent executing vector instructions. Since a vector instruction can last hundreds of cycles in executing, $M_v$ and $A_v$ can guide us in understanding how efficiently we are using the computational resources.
\item $\displaystyle C_v = {c_v}/{i_v}$ is the {\bf cycles per vector instruction}, or vector CPI, or vCPI. The interpretation of $C_v$ requires a bit of architectural knowledge to be understood. Assuming an ideal code only executing back to back FMA instructions we would have $C_v = 32$ on the RISC-V VEC and $C_v = 8$ on the NEC SX-Aurora. Those are in fact the latency of an arithmetic instruction on such architectures.
\item 
$ \overline{vl} = \frac{1}{i_v}\sum_{k=0}^{k<i_v}vl_{k}$
is the {\bf average vector length} of the vector instructions, or AVL, which is the average number of elements used by the vector instructions.
\item $\displaystyle E_v = \overline{vl}/vl_{\mbox{max}}$ is the {\bf vector occupancy}, a number between 0 and 1 that indicates how well vector instructions effectively leverage the maximum vector length given by the machine. $vl_{\mbox{max}} = 256$ double precision elements on NEC SX-Aurora and RISC-V VEC, while $vl_{\mbox{max}} = 8$ on Intel AVX512\footnote{This is not completely accurate, because the use of different vector lengths in Intel implies the use of different kind of instructions corresponding to different x86 SIMD extensions, but it serves for the formal definition taking as an example AVX512.}.

\end{itemize}
To gain a deeper understanding of the instructions executed, we considered an instruction hierarchy for later analysis purposes to classify each instruction into a type.
Figure \ref{figInstructionHierarchy} shows the instruction hierarchy tree.

The executed instructions are split into ``Scalar'', ``Vector'', and ``Vector configuration'' instructions.
``Vector configuration'' instructions configure the vector length and the element width of the next vector instructions.
The box called ``Vector'' represents the set of instructions executed on the vector processing unit.
Among vector instructions, we can find instructions that execute arithmetic operations, instructions to access the memory, and control lane instructions.
``Control lane'' vector instructions include all vector instructions that do not compute any arithmetic values nor communicate with memory, such as moves, shifts or sign extensions.

\subsection{Application}\label{secApplication}

%
We have chosen Alya, an advanced application running on most of the modern HPC architectures~\cite{banchelli2020benchmarking}, as the subject for analysis, porting, and efficient execution on the RISC-V vector prototype.
Alya is a computational mechanics application that solves complex coupled multi-physics problems written in Fortran~\cite{vazquez2016alya,garcia2020runtime}.
%
%
It is part of the Unified European Applications Benchmark Suite (UEABS)\footnote{\url{https://repository.prace-ri.eu/git/UEABS/ueabs}}, a set of HPC applications selected as representative, worth studying and maintaining into the future, and relevant for the procurement of next-generation computing systems.

The application is divided into 12 modules, each one solving different physics problems.
Nastin is the module that solves the computational fluid dynamics (CFD) simulations using the Navier-Stokes equations.
CFD applications are often structured into two primary operations:
{\em i)} matrix and Right Hand Side (RHS) assembly and
{\em ii)} algebraic linear solver

We isolate the matrix and right-hand side (RHS) assembly components, which constitute a computationally intensive section of the Nastin module. By encapsulating this specific segment of the code into a mini-application, we are able to investigate the implications associated with its execution on long-vector architectures.
Our methodology relies on compiler-introduced auto-vectorization.
The compiler identifies loops within the code and endeavors to vectorize the loop bodies. 
Consequently, we divided the mini-application into eight phases, with each phase encompassing one or multiple loops extracted from the assembly algorithm.
The division of these phases is artificial and does not align with algorithmic or physics-related logical steps. Its sole purpose is to ease the identification of vectorization hotspots, which we analyze using the metrics introduced in Section~\ref{secPerformanceMetrics}.
%
%
The corresponding operations of each phase are outlined as follows: 

\begin{itemize}
  \item Phase 1 and 2 gather data from the global data structures (at the mesh level) to the local ones (at the element level) used within the kernel. These two phases only include memory accesses and no floating-point operations are executed. This phase is candidate to be memory-bound limited.
  \item Phase 3 includes operations for calculating the Jacobian at the integration points using nested loops. It mostly involves floating-point operations and we can expect this phase to stress the floating point functional units of the VPU.
  \item Phase 4 implements several nested loops over integration points, nodes and space dimensions, to compute some intermediate arrays at the integration points. Similar to Phase 3, the inner loop involves mostly floating-point operations.
  \item Phase 5 is performing calculations to compute several elemental arrays needed by the time integration scheme. This phase involves several non-nested loops and floating-point operations are involved.
  \item Phase 6 implements several nested loops over integration points, nodes and space dimensions, to compute the convective term contribution to the elemental residuals of the momentum equations (high-hand sides). It is structured in three set of nested loops involving heavy arithmetic floating point operations.
  \item Phase 7 performs the calculation of the viscous term contribution to the elemental matrices and right-hand sides. Element matrices are computed only if the semi-implicit numerical scheme is considered. It includes arithmetic floating-point operations.
  \item Phase 8 checks for valid elements, and assembles contributions from local element arrays into the global vectors and matrices. It includes some arithmetic but it is dominated by memory accesses.
\end{itemize}

\vs is a compile-time configurable parameter of Alya which represents the amount of elements the kernel processes per single call from a bigger mesh.
In our work, we studied 6 \vs values: 16, 64, 128, 240%
\footnote{The value \vs =240 is the only one that is not power of 2. We consider this value for micro-architectural reasons: in the RISC-V VEC prototype performance are maximized when the vector length is a multiple of 8 (the number of functional units) and 5 (a micro-architectural parameter related to FSM implementation), thus vector length that are multiple of 40 are the ones that most likely maximize the performance.}%
, 256 and 512.

Note that the terms \vs and {\em vector length} look similar but have nothing to do with each other.
\vs is an application software parameter that affects the data structure of Alya.
The vector length, or VL, is the size of the registers on which the vector instructions dynamically operate.
It is an architectural feature closely tied to the vector architecture.

Alya supports different parallel programming models and is proven to scale up to thousand of cores~\cite{banchelli2020benchmarking}. However, since our work focuses on single-core performance, experiments are executed serially on the given HPC platforms.

\subsection{Comparative HPC platforms}\label{secHpcPlatforms}

{\bf \blind{MareNostrum4}} is the main supercomputing platform of \blind{the BSC ranked 29th in the Top500 (June 2019)}.
Each node houses 2 Intel Xeon Platinum 8160 sockets.
Intel Xeon Platinum features AVX-512, a SIMD extension implementing instructions of 512-bits wide, allowing computations of 8 double-precision elements.
The SIMD unit of a core of \blind{MareNostrum4} allows to reach a peak performance per core of 67.2~GFLOPS when issuing 2~FMAs instructions per cycles that operates 8 elements at a frequency of 2.1~GHz.
%


{\bf NEC SX-Aurora} is an accelerator developed by NEC which bases its computational power on a long-vector architecture.
Each core is able to operate vector registers of 16-kbits, \ie 256 double-precision elements, using 32 functional units working in parallel.
This means that a vector FMA instruction performs 512 FLOPS and needs 8 cycles to graduate.
Each SX-Aurora accelerator houses 8 Vector Extension (VE) cores consisting on one scalar unit coupled with a Vector Processing Unit (VPU).
Thus, each VPU has 96 lanes operating floating-point operations in parallel.
The peak double precision performance of a single VE is 307.2~GFLOPS.
%

Both these platforms are intended to be used as a reference point to compare the results with the other platforms used in this paper.
In Table \ref{tabHWConfiguration} we summarize the main software and hardware characteristics from the three platforms.
All the entries are measured per core.

\input{35-table-platforms.tex}

%% file: 35-table-platforms.tex
\begin{table}[htbp]
  \caption{HPC platforms: hardware and software configuration.}
  \label{tabHWConfiguration}
  \rowcolors{2}{gray!15}{white}
\resizebox{\columnwidth}{!}{%
\begin{tabular}{r|c|c|c}
    \rowcolor{gray!30}
                                                                      & \textbf{RISC-V VEC}                                           & \textbf{\blind{MareNostrum 4}} & \textbf{SX-Aurora} \\ \hline\hline
Architecture                                                          & \begin{tabular}[c]{@{}l@{}}RISC-V +\\ RVV v0.7.1\end{tabular} & Intel x86              & VE20B              \\ 
\begin{tabular}[r]{@{}r@{}}Cores per \\ socket\end{tabular}           & 1                                                             & 24                     & 8                  \\
\begin{tabular}[r]{@{}r@{}}Frequency\\ {[}MHz{]}\end{tabular}         & 50                                                            & 2100                   & 1600               \\
\begin{tabular}[r]{@{}r@{}}Bandwidth\\ {[}Bytes/cycle{]}\end{tabular} & 64                                                            & 11.20                  & 120                \\
\begin{tabular}[r]{@{}r@{}}Throughput\\ {[}FLOP/cycle{]}\end{tabular} & 16                                                            & 32                     & 192                \\ \hline\hline
Compiler                                                              & flang 18.0.0                                                  & ifort 2018.4           & nfort 5.0.2        \\
OS                                                                    & Ubuntu 21.04                                                  & Suse 12 SP2            & VEOS              \\
\end{tabular}%
}
\end{table}

%% file: 38-methodology.tex
\section{Methodology}\label{secMethodology}

%

In the context of this paper, co-design refers to the process of meticulously analyzing the performance of a scientific application on a system and identifying the factors that limit its performance from a holistic point of view.
We consider that we can act on three main fields:
{\em i)} the hardware, as executing the application on novel architectures can raise questions about the architecture itself and we are still on time to influence some design decisions;
{\em ii)} the system software, \eg the compilation of an application can reveal features that are not supported by the compiler;
{\em iii)} and the application, as sometimes, the application itself incorporates ``features'' that compromise portability, such as the use of intrinsics.

Since our goal is to co-design with complex scientific codes rather than benchmarks, we employ a flexible set of tools to instrument the code, collect execution traces, and analyze the application with varying levels of granularity. We repeatedly conduct this analysis to continuously identify which factors have the most significant impact on each phase under study. For instance, if we compile the application and detect an unvectorized loop due to a complex data structure, we address it and repeat the process to determine if any other limitations still affect the application's performance.
To achieve this, we utilize the tools introduced in Section~\ref{secPerformanceTools}. 

One might criticize this fine-tuning approach for a specific architecture, as it could hinder portability. However, our method takes this aspect into careful consideration. Therefore, we not only propose modifications to the application in collaboration with the developers but also ensure that all changes made to enhance vectorization do not penalize performance on other HPC systems.
As described in Section~\ref{secHpcPlatforms}, in our study, we validate our co-design efforts on a ``traditional'' HPC cluster based on Intel x86 CPUs and another vector accelerator that utilizes large vectors, following an approach similar to the one adopted by the RISC-V VEC.

%% file: 50-implementation.tex
\section{Implementation and optimization}\label{secImplementation}

In this section we present the mini-app vectorization process including the code optimizations and their respective performance analysis in RISC-V vector system described in Section~\ref{secRiscVProto}.

The first step is to instrument the mini-app so that we can separate phases, \ie regions of code, where critical loops are located.
%
%
Table \ref{tabPhaseWeightsScalar} shows the percentage of cycles spent in each phase when running the mini-app scalar  on the RISC-V vector system with vectorization disabled.
%

\begin{table}[htbp]
\centering
\caption{Percentage total cycles spent per phase.}
\label{tabPhaseWeightsScalar}
\includegraphics[width=\columnwidth]{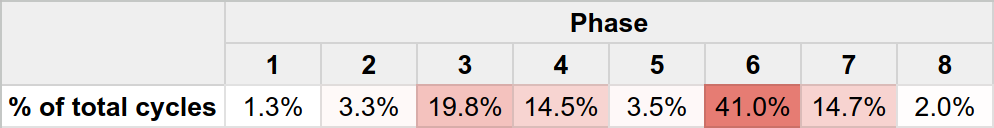}
\end{table}

The percentage of clock cycles is color coded from white to red being white the less consuming phase and red the most consuming one.
As we can see, the most consuming phases are 6, 7, 3 and 4 which account for the 90\% of total cycles.
%
%
%
%
After this scalar preliminary study, we compile and execute the mini-app enabling the auto-vectorization to check the implications when running with long-vector architectures.

\begin{figure}[htbp]
\centering
\includegraphics[width=\columnwidth]{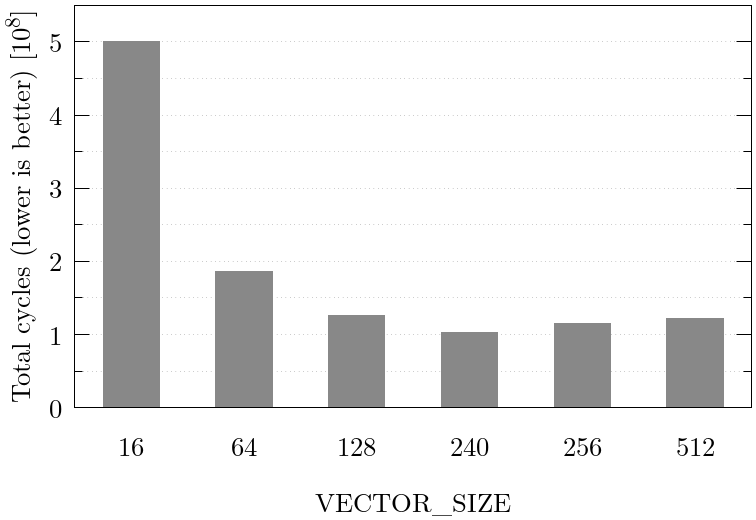}
\caption{Total cycles spent in the vanilla mini-app enabling auto-vectorization.}
\label{figVectorCycles}
\end{figure}

Figure \ref{figVectorCycles} shows the cycles executing the original mini-app after enabling the auto-vectorization:
$x$-axis represent each \vs tested and $y$-axis represent execution time in cycles (lower is better).

As we can observe, when enabling auto-vectorization, the \vs parameter has an important influence.
The fastest configuration is the one corresponding to \vs $ = 240$.

To quantify which phases are being auto-vectorized and how many vector instructions are present, we can observe Table~\ref{figVanillaInsMix}.
The heatmap in Table~\ref{figVanillaInsMix} shows the vector instruction mix.
Columns represent the phases and rows the values of the \vs parameter tested.
The colour intensity represents the vector instruction mix going from 0\% (red, no vector instructions) to 100\% (green, auto-vectorized code).

\begin{table}[htbp]
\centering
\caption{Vanilla vector instruction mix, $M_v$, \ie percentage of vector instructions wrt the number of total instructions executed.}
\label{figVanillaInsMix}
\includegraphics[width=\columnwidth]{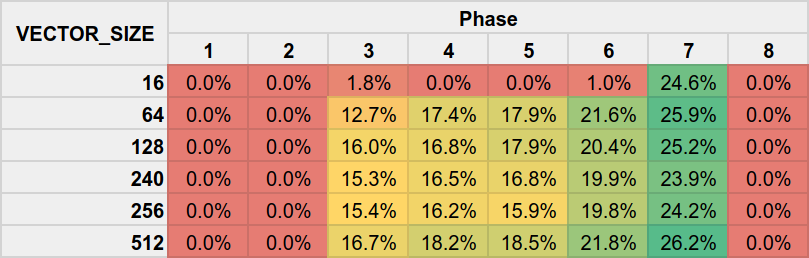}
\end{table}

Cells with values of vector instruction mix close to zero represent phases and values of \vs for which the compiler was unable to auto-vectorize, \eg phases 1, 2 and 8.
Interestingly, with \vs$={16}$ the compiler is only able to vectorize phase 7 (and very little phases 3 and 6).
Increasing the value of \vs, grows the ratio of vector instructions of phases 3 and 6.
Also, it helps the compiler to auto-vectorize phases 3 and 5.
Values of \vs $>64$ do not influence the ratio of vector instructions.

\begin{figure}[!htbp]
\centering
\includegraphics[width=\columnwidth]{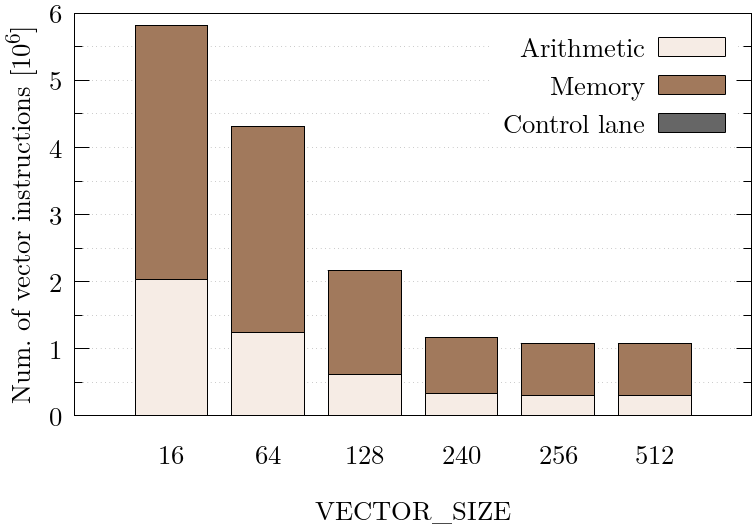}
\caption{Absolute number and type of vector instructions executed when enabling auto-vectorization.}
\label{figVanillaAbsVector}
\end{figure}

The heatmap in figure \ref{figVanillaInsMix} is useful to show the percentage of vector instructions, but hides the absolute number of vector instructions executed per each \vs.
This information is shown in Figure \ref{figVanillaAbsVector}, where we plot the absolute number of vector instructions.
Each colour represent a vector instruction type.
The $x$-axis represent each \vs tested and $y$-axis the number of vector instructions.

The reader should observe how the number of vector instructions executed decrease as the \vs parameter increases.
This is expected, because with a higher \vs the data structures of the application can make use of more elements inside each vector instruction, thus reducing the total number of instructions.
Also note that throughout execution, the VPU is only performing arithmetic operations or moving data between vector registers and memory, as there are no control lane vector instructions.
Additionally, almost 70\% of vector instructions are memory type.
This high percentage of memory accesses causes the mini-app not to take fully advantage of the computing power of the VPU.

Table \ref{tabPhase6InitialStudy} shows the vCPI, the AVL of vector instructions and the total number of vector instructions per tested \vs in phase 6, as it is the most time-consuming phase, where almost all the floating-point operations reside. 

\begin{table}[htbp]
  \centering
  \rowcolors{2}{gray!15}{white}
  \caption{vCPI, AVL and number of vector instructions in phase 6.}
  \label{tabPhase6InitialStudy}
  \begin{tabular}[\textwidth]{cccc}
    \rowcolor{gray!30}
    \vs & vCPI & AVL & Number vector instructions \\
    16 & 9.71 & 16 & \text{14.3$\times$10\textsuperscript{5}} \\
    64 & 23.39 & 64 & \text{19.1$\times$10\textsuperscript{5}} \\
    128 & 28.56 & 128 & \text{9.6$\times$10\textsuperscript{5}} \\
    240 & 41.19 & 240 & \text{5.1$\times$10\textsuperscript{5}} \\
    256 & 43.10 & 256 & \text{4.7$\times$10\textsuperscript{5}} \\
    512 & 45.30 & 256 & \text{4.7$\times$10\textsuperscript{5}} \\
 \end{tabular}
\end{table}

As we can see, increasing \vs increments the AVL of the vector instructions, thus processing more elements per single vector instruction. 
This is good for the overall performance as it uses more efficiently the resources of the VPU.
Also, this affects the vCPI, increasing the latencies of the vector instructions as more cycles are required to complete each vector instruction.
In this phase, the VPU responds with vector instructions with a vector length equal to \vs$_{i}$ because the inner-most loops have \vs parameter as induction variable.

The reader should observe that the run with \vs$={256}$ has a vCPI > 32 cycles.
We measured the vector floating-point Fused Multiply-Add (FMA) instruction latencies through a synthetic benchmark and found that one vector FMA takes around 32 cycles with a vector length of 256, while with a lower vector length takes less cycles.
We conclude that the vector memory and arithmetic pipelines are not fully overlapped, but we are not that far from achieving the ideal case where the vector FMA latency is near 32 cycles.

Observing the total number of vector instructions, we see a inverse proportional reduction with the AVL, because the more elements we process per vector instructions, the fewer vector instructions we have to execute.

Note that the vCPI does not increase linearly with the vector length.
For example, jumping from vector length 64 to 128 results in a 2$\times$ decrease in vector instructions, but only in a 1.22$\times$ slowdown in the vCPI.
An efficient exploitation of vector architectures often relies on increasing the average vector length of vector instructions to the $vl_{\mbox{max}}$, as we execute wide vector instructions that process many elements in a row, thus avoiding the overhead of the cycles in the early stages of the pipeline.

After this initial evaluation of the original mini-app with vectors, we calculated the percentage of total cycles spent in each phase, taking into account the \vs parameter, since we have seen that this parameter has a strong impact when the mini-app uses vector instructions.

\begin{figure}[!htbp]
\centering
\includegraphics[width=\columnwidth]{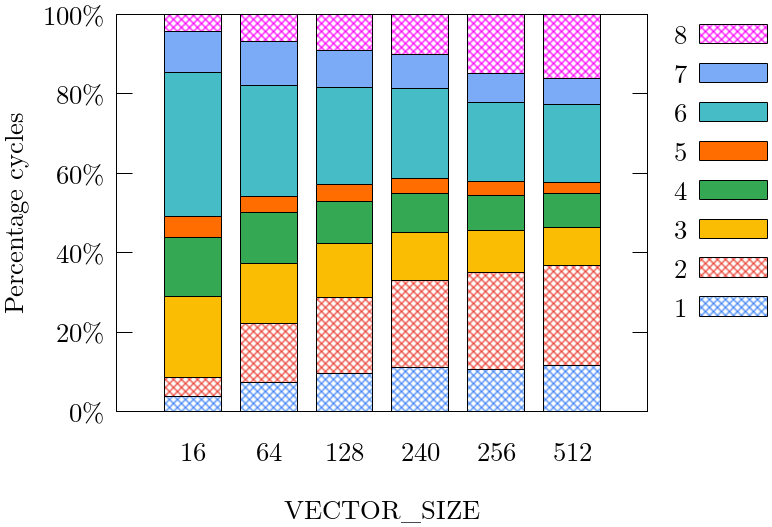}
\caption{Percentage cycles spent per phase.}
\label{figStackedPercVanilla}
\end{figure}

Figure \ref{figStackedPercVanilla} represents the percentage of cycles spent per phase.
Each colour represents a phase.
$x$-axis represent the \vs tested and $y$-axis mean percentage of total cycles.
The most cycle-intensive phases which used to consume around 90\% of time, mentioned in Table~\ref{tabPhaseWeightsScalar}, now consume approximately 50\% of cycles, caused by the highly vectorization of these phases.
We can also observe that non-vectorized phases, marked with a custom pattern, consume a greater percentage of cycles.
For example, phase 1 and phase 2 which used to cost 4\% of the execution, see table \ref{tabPhaseWeightsScalar}, now have a percentage around 38\%, as we increase the value of the \vs parameter.

 
In Figure \ref{figStackedPercVanilla} we show that the most time-consuming phase is phase 2, thus we focus on enable efficient auto-vectorization in phase 2, listed in Algorithm~\ref{lst:original-p2}.
None of these loops compute any floating-point operation, only move data between the general structures and those specifically used in the mini-app.
Leaving aside the fact that it does not compute any useful operation, data movement operations can be expensive, usually when the memory access is strided or indexed, and can be improved by leveraging vector memory instructions.

\begin{lstlisting}[language=Fortran, caption={Original code phase 2.},label={lst:original-p2}]
do ivect = 1, VECTOR_DIM                      
   do inode = 1, pnode
     !! WORK
   end do
end do
\end{lstlisting}
\begin{lstlisting}[language=Fortran, caption={Final code phase 2.},label={lst:final-p2}]
do inode = 1, pnode
   do ivect = 1, VECTOR_DIM
     !! WORK
   end do
end do
\end{lstlisting}

\begin{figure}[!htbp]
\centering
\includegraphics[width=\columnwidth]{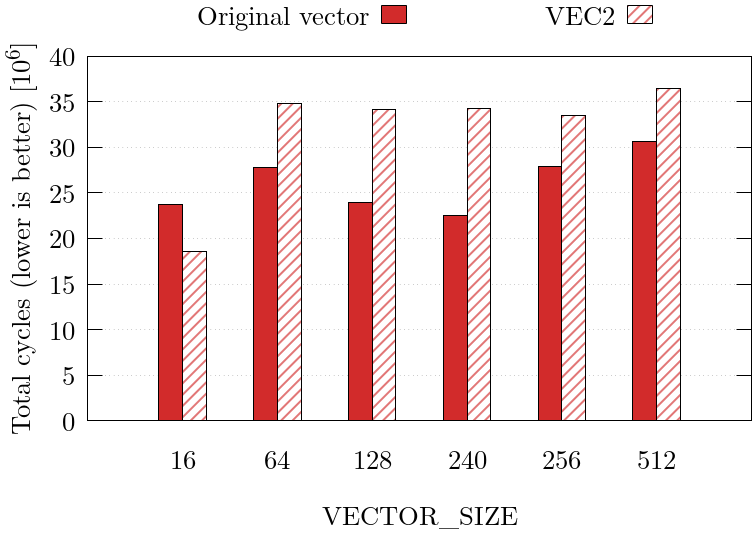}
\caption{Absolute cycles phase 2.}
\label{figVEC2}
\end{figure}

After inspecting the generated assembly and using the LLVM vectorization remarks, we detect that the compiler is fetching, from memory, the {\tt VECTOR\_DIM} parameter each iteration.
Knowing that {\tt VECTOR\_DIM} dummy argument is equal to \vs, we hide to the compiler information known at compile-time.
Finally, deleting {\tt VECTOR\_DIM} dummy argument from the mini-app declaration and creating a constant variable within the mini-app known at compile-time, enables the compiler to auto-vectorize phase 2.
After this optimization, phase 2 data movement operations use the VPU, instead of the scalar core as before.
From now on we will name this optimization VEC2.
Figure \ref{figVEC2} shows the cycles spent in phase 2.
Different bars represent different optimizations.
$x$-axis represent the \vs tested and $y$-axis cycles.

As we can observe, enabling auto-vectorization of phase~2 has been counter-productive and degraded the performance with respect to the original version.
The cycles spent in phase~2 have risen significantly for all \vs except for \vs$={16}$.

We leveraged the RISC-V emulator to measure the AVL of the vector instructions and therefore explain the performance drop within phase~2.
The measured AVL of the vector instructions is 4 double-precision elements (out of the 256 double-precision elements available on the vector register).
Decoding, issuing and dispatching vector instructions to the VPU computing only 4 elements produces significant overhead to the overall execution.

To fix this low AVL we re-structure the loop induction variables to force {\tt VECTOR\_DIM} to be the inner-most loop. This allows the compiler that autovectorize the inner-most loop to ask the hardware for larger vector length, thus making a more efficient use of the VPU instructions.
We name this optimization IVEC2.
Algorithm~\ref{lst:final-p2} shows the resulting loop of phase~2.

\begin{figure}[!htbp]
\centering
\includegraphics[width=\columnwidth]{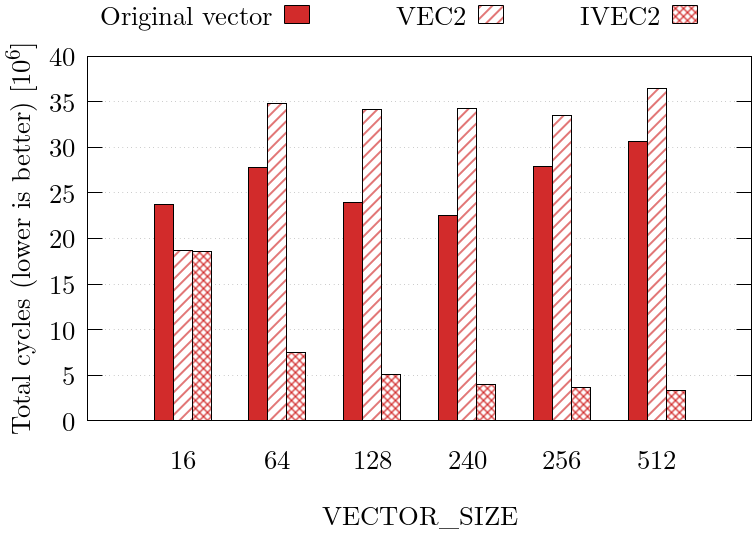}
\caption{Resulting cycles phase 2.}
\label{figIVEC2}
\end{figure}

Figure \ref{figIVEC2} shows the resulting performance of phase~2.
Different bars represent different optimizations.
$x$-axis represent the tested \vs and $y$-axis absolute cycles.

Re-structure the loop induction variables, forcing {\tt VECTOR\_DIM} to be the inner-most loop induction variable, results in a significant speed-up up to 7.38$\times$ with respect to the original version with \vs$={256}$.
This optimization produces vector instructions with ``vector length'' = \vs.
Increasing \vs results in a higher speed-up, as we leverage the benefits of increasing vector occupancy, even if all vector instructions are memory type (\ie not computing any floating-point value).

The next optimization targets phase~1, another phase which the compiler is unable to auto-vectorize.
Algorithm~\ref{lst:phase-p1} shows the pseudo-code of phase~1.

\begin{lstlisting}[language=Fortran, caption={Original code phase 1.},label={lst:phase-p1}]
do ivect = 1, VECTOR_DIM
     !! WORK A
     !! WORK B
end do
\end{lstlisting}
\begin{lstlisting}[language=Fortran, caption={Final code phase 1.},label={lst:final-phase-p1}]
do ivect = 1, VECTOR_DIM
     !! WORK A running without vector instructions
end do
do ivect = 1, VECTOR_DIM
     !! WORK B running vector instructions
end do
\end{lstlisting}

LLVM vector remarks showcased that loops inside work B are being auto-vectorized and loops inside work A are not auto-vectorized.
Leveraging Vehave emulator, we checked that work B was executed scalar, even though the compiler inserted vector instructions. 
Since they are inside the same outer loop, during runtime, it decides to opt for scalar execution, as one of outer loop parts is not vectorizable.
Splitting them into two different outer loops, like Algorithm~\ref{lst:final-phase-p1}, allows work B to use vector instructions.
We name this optimization VEC1.

\begin{figure}[!htbp]
\centering
\includegraphics[width=\columnwidth]{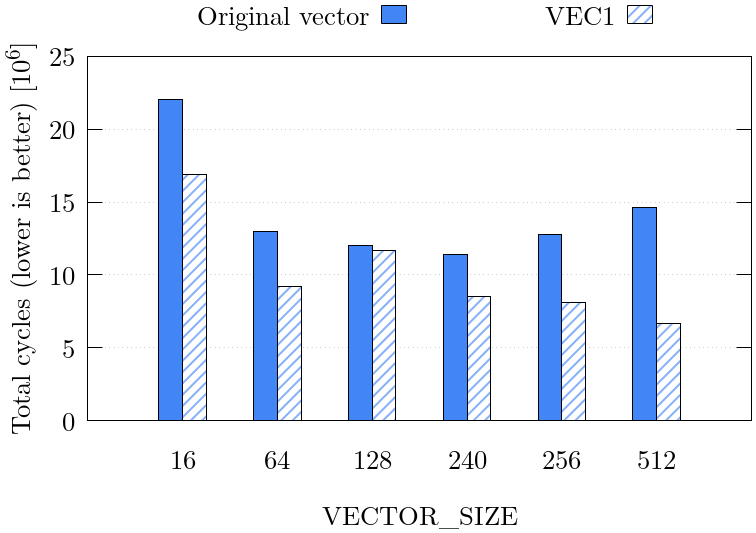}
\caption{Resulting cycles phase 1.}
\label{figVEC1}
\end{figure}

Figure \ref{figVEC1} shows the cycles spent in phase~1.
Different bars represent different optimizations.
$x$-axis represent the \vs tested and $y$-axis absolute cycles.

As we can see, this optimization does not have the same impact as IVEC2, as we are not running the whole phase with vectors.
We are able to achieve a speed-up of 2$\times$ in configuration with \vs$={512}$.
In other configurations, the optimization generates speed-ups ranging between 1.03$\times$ to 1.56$\times$ with respect to the original version.
Increasing \vs parameter does not scale the performance as with IVEC2 optimization.
A possible approach to increase the speed-up could be to further investigate how to vectorize the whole phase, since we are only using vector instructions in work~B.

\begin{figure}[!htbp]
\centering
\includegraphics[width=\columnwidth]{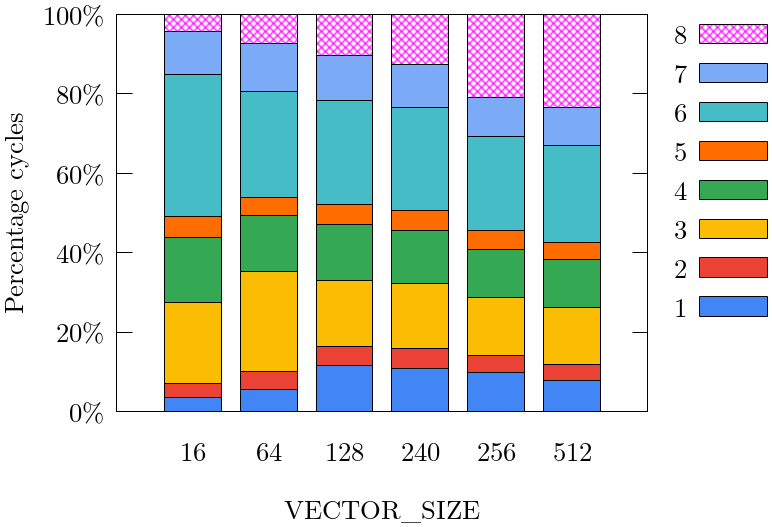}
\caption{Percentage total cycles spent per phase after optimizations.}
\label{figStackedPercFinal}
\end{figure}

Figure \ref{figStackedPercFinal} shows the resulting percentage of cycles spent per phase after applying the optimizations proposed to achieve a higher vectorization and use the full vector length capacity.
As we can observe, the optimizations in phase 1 and phase 2 resulted in a narrow percentage of cycles, with respect to the original version seen in Figure \ref{figStackedPercVanilla}.
The non-vectorized phase 8, marked with a custom pattern, keeps increasing as we increment the \vs parameter, while the other phases remain almost constant when running with \vs~$\geq 128$. 

%% file: 60-evaluation.tex
\section{Evaluation}\label{secEvaluation}

In this section we present a general evaluation of the mini-app, including the implications when increasing \vs and the speed-ups resulting from the optimizations.
Finally, we demonstrate the portability of our optimizations on another vector architecture, NEC SX-Aurora, and the HPC cluster \blind{\mn}.
This evaluation takes into account all the optimizations previously proposed.

\begin{figure}[htpb]
  \centering
  \includegraphics[width=\columnwidth]{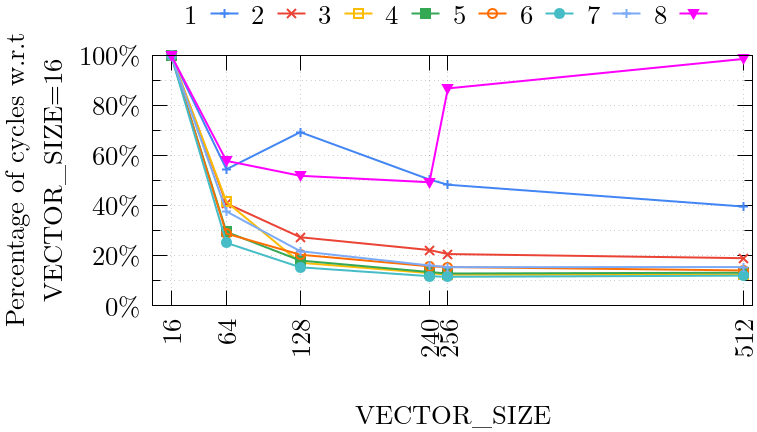}
  \caption{Percentage of cycles w.r.t \vs$={16}$ (lower is better).}
  \label{fig:cycle-scalability-final}
\end{figure}

Figure \ref{fig:cycle-scalability-final} shows the percentage of cycles with respect to \vs$={16}$ configuration in the RISC-V vector prototype, useful for quantifying performance improvements between 0\% and 100\%. 
$x$-axis represent the tested \vs and $y$-axis show the percentage of cycles relative to \vs$={16}$.
Each line represent a phase.

Experimental data tells us that percentages $\ge 100\%$, means generally no or serious problems in the vectorization, while percentages $> 30\%$ may highlight some problem of vectorization, percentages $\le 30\%$ generally indicates a good sign for vectorization.

On the one hand, highly auto-vectorized phases exhibit a reduction in the percentage of cycles with respect to \vs${=16}$ when increasing \vs, reaching 20\%.
On the other hand, phase 1 and phase 8 deviate from the desired behaviour. We will evaluate these two phases independently further in this section.
For now on, we focus on the ones that present a decrease in the percentage of cycles with respect to \vs${=16}$.

\begin{figure}[htpb]
  \centering
  \includegraphics[width=\columnwidth]{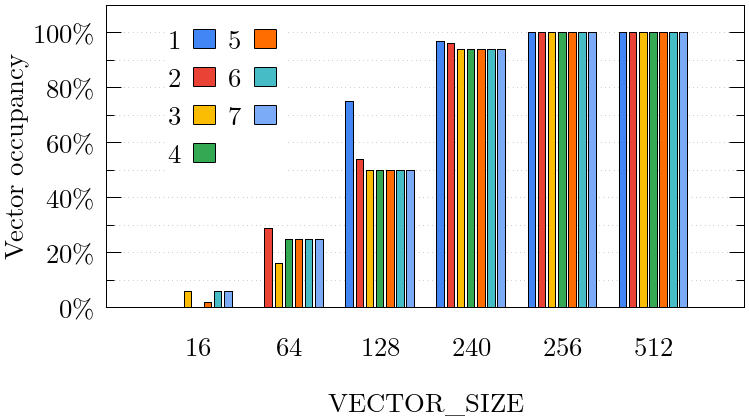}
  \caption{Vector occupancy (higher the better).}
  \label{fig:vector-efficiency}
\end{figure}

To justify these phases, we can rely on the vector occupancy to understand how well the vector registers are used compared to their maximum physical size allowed by the architecture.
Figure \ref{fig:vector-efficiency} shows the vector occupancy.
$x$-axis represent the tested \vs and $y$-axis the percentage of vector occupancy.
Each bar represent a phase.

Indeed, the vector occupancy reaches near 100\% in configurations with \vs close to the maximum physical size allowed by the architecture (256 double-precision floating-point).
Since phase 8 is not vectorized and therefore has no vector occupancy, we have omitted to plot the bar for clarity.

To understand the cycle percentage with respect to \vs$={16}$ of poorly and non-vectorized phases, we speculated about memory hierarchy overheads.
The curves depicted in Figure \ref{fig:cycle-scalability-final} for phase 1 and phase 8 are similar to a mixture of curvatures from L1 data cache misses per kilo-instruction and the percentage of memory instructions.
%
We can prove this hypothesis using multiple linear regression, often used to quantify the relationship between a dependent variable and multiple independent variables.
The closer the coefficient of determination is to 1, the better the results can be explained by the model.
Table \ref{tab:coefficient-determination} shows the resulting coefficient of determination using as dependent variable the number of cycles of phase 1 and phase 8, and as independent variables the number of L1 data cache misses (DCM) per kilo-instruction and the percentage of memory instructions.

\begin{table}[htbp]
  \centering
  \caption{Coefficient of determination phase 1 and phase 8.}
  \label{tab:coefficient-determination}
  \rowcolors{2}{gray!15}{white}
  \begin{tabular}[\textwidth]{cc}
    \rowcolor{gray!30}
    Phase & CoD (R\textsuperscript{2}) \\
    Phase 1 & 0.903 \\
    Phase 8 & 0.966 \\
  \end{tabular}%
\end{table}

Finally, after evaluate the performance implications of the \vs parameter, we proceed to analyze the performance of each optimization with respect to scalar execution with \vs$={16}$.

\begin{figure}[htpb]
  \centering
  \includegraphics[width=8cm]{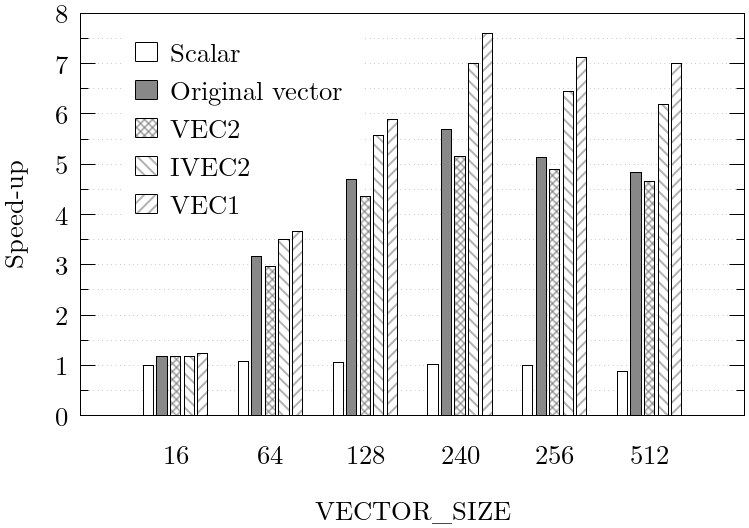}
  \caption{Speed-up with respect to scalar \vs$={16}$.}
  \label{fig:speedup-global}
\end{figure}

Figure \ref{fig:speedup-global} shows the speedup of the mini-app with respect to \vs$={16}$.
$x$-axis represent the tested \vs.
$y$-axis means speed-up.
Each bar represents an optimization.
Optimizations are accumulative in the same way they are presented in the section.

The original mini-app auto-vectorization achieves speed-ups between 3$\times$ and 6$\times$, this demonstrates the excellent quality of the LLVM-based \blind{EPI} compiler.
The speed-up increments as \vs parameter increases up to \vs$={240}$, then the speed-up decreases and remains stable.
The VEC2 optimization yields to negative performance results as it was expected, compared to the original version. However, it is a step towards IVEC2 optimization which overpasses original version speed-up across all \vs and follows the same pattern where \vs$={240}$ is the faster configuration.
The VEC1 optimization it's not that noticeable, like IVEC2, although reaches speedups between 3.5$\times$ and 7.6$\times$.

The reason why \vs$={240}$ configuration is faster than the others, even though it has a lower vector occupancy, is because it has been quantified that vector floating-point instructions with vector length equal to 240 elements wide have a higher throughput of elements calculated per cycle than the other ones, due to the way the state machine is defined.
This effect causes this configuration to outperform the other ones.

We conclude the RISC-V vector performance evaluation stating that \vs$={240}$ configuration is the faster configuration running in the Vector Processing Unit (VPU), achieving a speed-up of 7.6$\times$ over the scalar execution.
As the VPU contains 8 lanes operating in parallel, the ideal speed-up we would obtain over the scalar execution is 8$\times$.
Hence, the analysis and optimizations proposed in our work are close to reach the peak performance on the RISC-V vector prototype.

\begin{figure}[htpb]
  \centering
  \includegraphics[width=8cm]{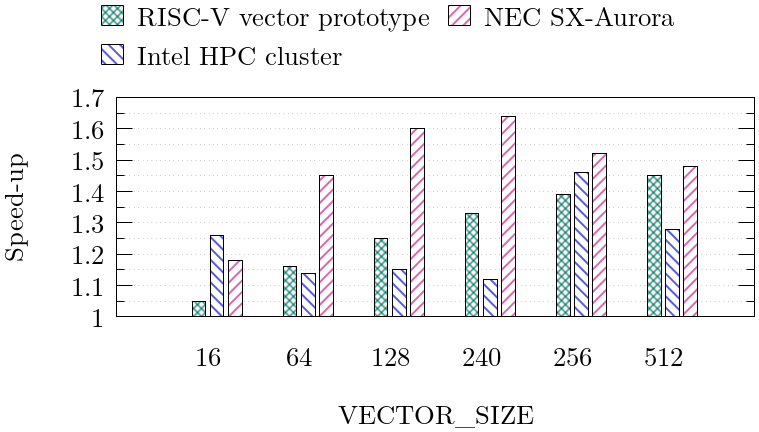}
  \caption{Speed-up optimizations on different HPC platforms.}
  \label{fig:speedup-platforms}
\end{figure}

Finally, we evaluate the optimizations in the different architectures used in this work.
Figure \ref{fig:speedup-platforms} shows the speed-up of the final optimizations to the mini-app with respect to the original vectorized implementation.
$x$-axis represent tested \vs and $y$-axis speed-up.
Each color represent a different platform.

\begin{figure}[htpb]
  \centering
  \includegraphics[width=8cm]{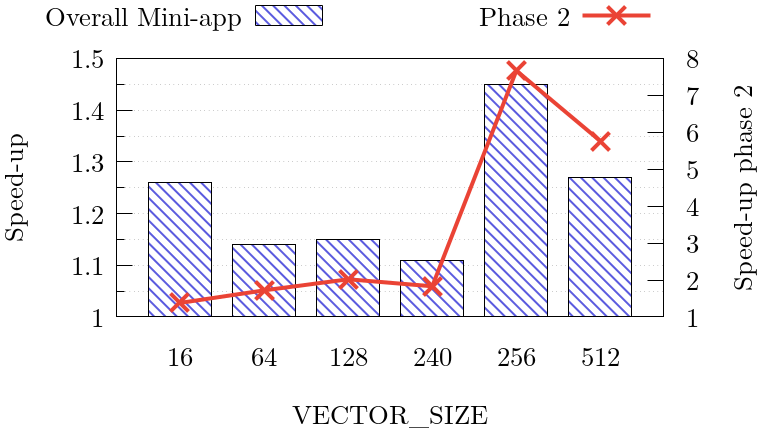}
  \caption{Speed-up optimizations on \blind{\mn}.}
  \label{fig:speedup-mn4}
\end{figure}

We observe that the enhancements proposed in this study are applicable to the diversity of the HPC platforms used.
The speed-up increases as \vs parameter increments, up to a 1.45$\times$ on the RISC-V vector prototype.
NEC SX-Aurora behave the same way as the RISC-V vector prototype until \vs$={256}$, then, the speed-up decreases.
Increment the \vs increases the cost of non-vectorized phases, such as phase 8, probably due to the complexity of indexed memory accesses.
Even though we are making better use of the vector instructions on the NEC platform, the weight of phase 8 outweighs the speed-up provided by the vector instructions, thus, reducing the enhancements.
We obtain a 1.64$\times$ speed-up in NEC SX-Aurora with \vs$={240}$.
To explain the \blind{\mn} speed-up, we must rely on phase 2 performance.
Figure \ref{fig:speedup-mn4} shows the mini-app overall speed-up and phase 2 speed-up.
$x$-axis represent the tested \vs, left $y$-axis the mini-app speed-up and right $y$-axis phase 2 speed-up.

As we can see, the influence of phase 2 impacts the overall mini-app speed-up.
This effect is caused by a decrease on the total number of L1 and L2 data cache misses and a reduction on the total instructions executed.

%% file: 20-related-work.tex
\section{Related Work}\label{secRelatedWork}

Concerning vector architecture based on RISC-V,
Cavalcante et al. in~\cite{cavalcante2023spatz} present a deep analysis of a RISC-V vector test chip that has been taped out at 12~nm. They study the energy consumption of the core+VPU as a function of the vector length. However, they focus on benchmarks rather than on scientific applications. We consider an added value to be able to perform early experiments with complex scientific codes on new architectures.
Perotti et al. in~\cite{perotti2022new} present an open-source implementation of a RISC-V core coupled with a VPU supporting the RISC-V vector extension 1.0. While the architecture is similar to the one used in our study, they focus more on the evaluation of the area of the chip and do not consider complex HPC applications for the evaluation.

Concerning CFD on recent HPC clusters and known vector architectures,
Oyarzun et al. in~\cite{oyarzun2018efficient} present a study of a CFD code on an Arm-based system: they provide insights and co-design feedback of running on early-days Arm-based prototypes. However the did not focus specifically on vectorization.
Bartels et al. in~\cite{bartels2002computational} provide an exhaustive review of the CFD approaches towards vector computation as well as a solid evaluation. However, the work focuses to the porting and optimization on existing platforms more than on providing co-design feedback.

Concerning co-design feedbacks in HPC,
Parkere et al. in~\cite{parkere2013role} provide an extensive review of the first co-design attempt made by the DoE for preparing codes to run efficiently on first Exascale systems. The approach proposed is to generate a set of benchmarks, mini- and proxy-apps that mimic the behaviour of real HPC applications. While this is the common approach to co-design still in place nowadays, we think our approach leaves more room for improvement because can be applied in an early stage of development of the new hardware and it allows to provide feedbacks, not only to the application developers, but also to the system software developers and system architects.

Sato et al. in~\cite{sato2020co} show the co-design process performed by the Japanese scientific community to have a set of HPC codes efficiently ported to the Arm-based Fugaku supercomputer. This work is extremely valuable, however, Banchelli et al. in~\cite{banchelli2021cluster} seems to highlight that the co-design effort for the Fugaku cluster does not look trivially portable to other HPC applications on the same cluster. We are confident that our method works for several applications and we proved that the performance and code portability are safe.

%% file: 70-conclusions.tex
\section{Conclusions}\label{secConclusions}

In this work, we optimize a kernel of a production CFD code to efficiently use a long vector unit while keeping the code portable and leveraging the auto-vectorization capabilities of the compiler.
The conclusions achieved by our study are a collection of lessons learned useful for HPC developers aiming to achieve performance portability of their codes across vector architectures.

%
We apply an iterative method to implement a fast co-design cycle:
we start by compiling the application with the auto-vectorizer offered by the LLVM compiler;
we identify the phases in which the compiler could not auto-vectorize (or the vectorization is suboptimal);
we identify what prevents efficient auto-vectorization and refactor the code to solve it;
we repeat the analysis to 
{\em i)} evaluate the impact of the code refactor {\em ii)} if successful, identify the next phase to optimize.

Gathering all the code refactors, we show 
up to 7.6$\times$ performance speedup when running in the target platform compared to the scalar execution and
up to 1.3$\times$ comparing with the code auto-vectorized by the compiler (with no code modifications).

We consider the Alya complex CFD code, focusing on a representative mini-app extracted from it, which serves as an illustration of one of its key computational kernels.
The mini-app is instrumented to facilitate the identification of different phases during analysis.
Throughout our study, we present techniques to enhance the number of vector instructions.
However, maximizing the number of vector instructions is not the sole ultimate objective;
it is crucial to ensure the optimal utilization of the entire vector length provided by the architecture.

This process is made feasible by leveraging the tracing and analysis tools available on the RISC-V vector prototype, supported by a multidisciplinary team.
The performance analyst identifies bottlenecks and collaborates with the compiler expert to pinpoint code sections not meeting desired performance standards.
Once identified, the compiler expert assesses whether the solution lies in enhancing the capabilities of the compiler or necessitates a code refactor.
If a code refactor is necessary, discussions with the code developer ensure the proposed changes are realistic and achievable.

While the process of porting an HPC application to a new architecture is often perceived as ``making the application machine-specific,'' we demonstrate that our code modifications ensure portability and yield performance improvements on other HPC platforms.
We validate our approach on a system with a large vector engine, NEC SX-Aurora, and a standard HPC cluster equipped with Intel CPUs, \mn.
This evaluation highlights that not only does performance portability hold, but it even improves in several cases.

We conclude our work highlighting detailed lessons learned for each profile of experts involved in the co-design cycle:

\textbf{Scientists / HPC application developers} -- 
When possible, provide loop limits at compile time instead of not constant variables: this allows the compiler to know the limits of the loops and vectorize them. 
When there are several nested loops, try to use the longer dimension in the most inner loop: this provides the compiler more margin to use longer vectors in the vector instructions.
Splitting loops into smaller units of work may aid the compiler in vectorizing easier certain sections of the code. Nevertheless, it is important to approach this suggestion with caution. In certain situations, it can be advantageous, but in others, particularly when multiple loops operate on the same data, it may be more beneficial to keep them together.

\textbf{System software developers} --
Since the LLVM Fortran compiler is under development, we identified and solved several unforeseen bugs.
These issues were documented and raised to the open-source community.
Additionally, we also revealed a critical issue in our infrastructure, highlighting disparities in auto-vectorization results between cross-compilation and native compilation processes.

\textbf{Hardware architects} --
Our findings indicate that the RISC-V VEC system achieves higher performance when operating with a vector length equal to 240 DP-elements rather than utilizing its full vector capacity (256 DP-elements). While this observation has been noticed in synthetic benchmark evaluations, our study quantifies its relevance applied to a production CFD code supported by profiling results.
The team was able to provide this feedback to the hardware team designing the RISC-V VEC system, encouraging addressing this micro-architectural insight in future RISC-V VEC prototypes.